\documentclass[sigconf]{acmart}

\AtBeginDocument{%
  \providecommand\BibTeX{{%
    \normalfont B\kern-0.5em{\scshape i\kern-0.25em b}\kern-0.8em\TeX}}}

\copyrightyear{2024} 
\acmYear{2024} 
\setcopyright{rightsretained} 
\acmConference[AHs 2024]{The Augmented Humans International Conference}{April 4--6, 2024}{Melbourne, VIC, Australia}
\acmBooktitle{The Augmented Humans International Conference (AHs 2024), April 4--6, 2024, Melbourne, VIC, Australia}
\acmDOI{10.1145/3652920.3652923}
\acmISBN{979-8-4007-0980-7/24/04}

\usepackage[hypcap=false]{caption}
\usepackage{subcaption}
\usepackage{amsmath}
\usepackage[capitalise]{cleveref}
\usepackage[font=small,skip=5pt]{caption}
\begin{document}


\title{iFace: Hand-Over-Face Gesture Recognition Leveraging Impedance Sensing}

\author{Mengxi Liu}
\email{mengxi.liu@dfki.de}
\affiliation{%
  \institution{German Research Center for Artificial Intelligence(DFKI)}
  \city{Kaiserslautern}
  \country{Germany}
  \postcode{67663}
}

\author{Hymalai Bello}
\email{hymalai.bello@dfki.de}
\affiliation{%
  \institution{German Research Center for Artificial Intelligence(DFKI)}
  \city{Kaiserslautern}
  \country{Germany}
  \postcode{67663}
}

\author{Bo Zhou}
\email{bo.zhou@dfki.de}
\affiliation{%
  \institution{German Research Center for Artificial Intelligence(DFKI)}
  \city{Kaiserslautern}
  \country{Germany}
  \postcode{67663}
}

\author{Paul Lukowicz}
\email{paul.lukowicz@dfki.de}
\affiliation{%
  \institution{German Research Center for Artificial Intelligence(DFKI)}
  \city{Kaiserslautern}
  \country{Germany}
  \postcode{67663}
}

\author{Jakob Karolus}
\email{jakob.Karolus@dfki.de}
\affiliation{%
  \institution{German Research Center for Artificial Intelligence(DFKI)}
  \city{Kaiserslautern}
  \country{Germany}
  \postcode{67663}
}

\renewcommand{\shortauthors}{Mengxi, et al.}

\begin{abstract}
Hand-over-face gestures can provide important implicit interactions during conversations, such as frustration or excitement. However, in situations where interlocutors are not visible, such as phone calls or textual communication, the potential meaning contained in the hand-over-face gestures is lost.
In this work, we present iFace, an unobtrusive, wearable impedance-sensing solution for recognizing different hand-over-face gestures. 
In contrast to most existing works, iFace does not require the placement of sensors on the user's face or hands. 
Instead, we proposed a novel sensing configuration, the shoulders, which remains invisible to both the user and outside observers. 
The system can monitor the shoulder-to-shoulder impedance variation caused by gestures through electrodes attached to each shoulder.
We evaluated iFace in a user study with eight participants, collecting six kinds of hand-over-face gestures with different meanings. 
Using a convolutional neural network and a user-dependent classification, iFace reaches 82.58 \% macro F1 score. 
We discuss potential application scenarios of iFace as an implicit interaction interface.

\end{abstract}



\begin{teaserfigure}
  \includegraphics[width=\textwidth]{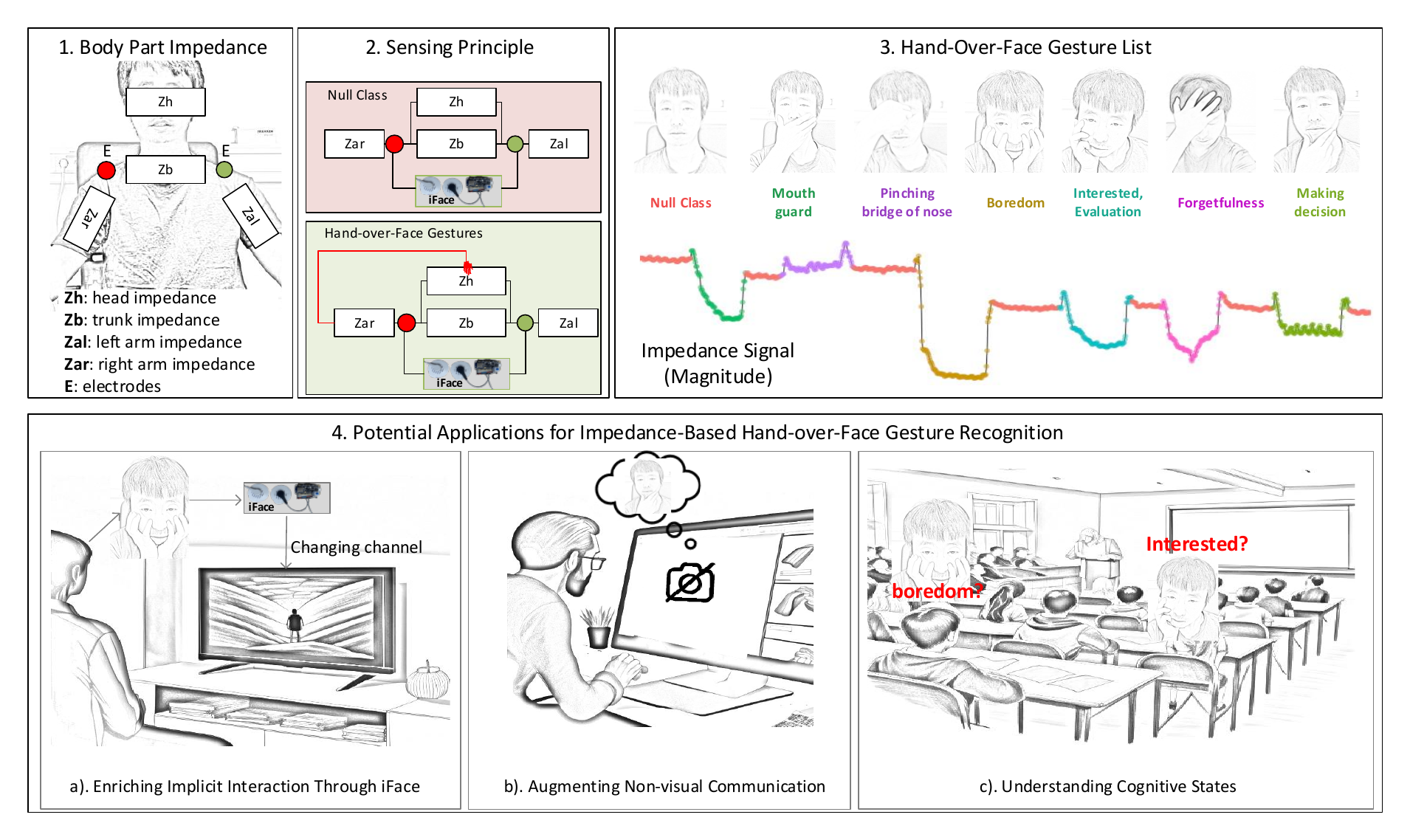}
  \caption{Schematic diagram and potential application scenarios of iFace. 
    \textbf{1} and \textbf{2}: the sensing principle of iFace is based on the body impedance variation caused by hand-face contact. 
    \textbf{3}: the impedance signal variation depends on the contact position and area when performing hand-over-face gestures.
    \textbf{4a}: iFace enabling implicit interaction between users and their TV, e.g. the channel changes automatically when boredom is detected. 
    \textbf{4b}: hand-over-face gestures recognized by iFace during an online meeting can replace low-quality camera images to convey bodily cues in real-time. 
    \textbf{4c}: iFace as an additional sensing modality for multi-modal inference of a user’s cognitive state in teaching scenarios.}
  \label{fig:teaser}
\end{teaserfigure}


\maketitle

\section{Introduction}

Implicit interactions, which convey sensations such as frustration or excitement, are manifested and encoded through verbal and nonverbal means, including voice, facial expressions, and body language \cite{naik2020improved}.
For example, people often hold their hands near their faces, forming gestures in conversations as a kind of nonverbal body language.
Spontaneous self-touches or self-grooming gestures are believed to be related to formulating thoughts, information processing, and emotion regulation \cite{freedman1977hands}.
Empirical evidence substantiates that certain hand-over-face gestures function as cues facilitating the recognition of cognitive mental states \cite{mahmoud2016automatic}.
However, in situations where interlocutors are not visible, such as phone calls or textual communication, the potential meaning contained in the hand-over-face gestures is lost.
Therefore, such kinds of non-visible communication ways suffer from a lack of contextual and emotional awareness \cite{hassib2017heartchat}.
Although emoticons are widely used to explicitly express emotion by users, misunderstanding of the emoticons still happens between users frequently \cite{lu2016learning}. 

Facial expressions and hand gesture recognition have been widely studied.
Most of the work is based on computer vision technology and deep neural networks, achieving remarkable accuracy \cite{hasani2017facial,li2020deep}. 
Yet, those approaches often suffer from privacy invasion, which limits their application on private and sensitive occasions.
One solution that respects privacy is to use the user's avatar, as in Meta Horizon, which requires a headset (e.g., Oculus) to be worn by the user. 
Which is bulky, not ubiquitous, and uncomfortable to wear daily. 
Here, wearable sensor-based approaches for facial expression and hand gesture recognition can offer a privacy-friendly and ubiquitous alternative.
However, most wearable solutions often require the sensors to be placed near the face or hands \cite{inmyface,samadiani2019review, li2019fusion}, which is potentially uncomfortable and has low social acceptance.
Contrarily, impedance-based sensing allows the electrodes to be placed on covered body parts and enables arbitrary gesture detection by measuring the impedance information between the two electrodes. The potential of impedance-sensing-based solutions for human activity recognition has been demonstrated in several works \cite{liu2024imove, liu2023ieat}.

In this work, we present iFace, an unobtrusive, wearable impedance-sensing device for hand-over-face gesture recognition.
In contrast to most existing works, iFace does not require the placement of sensors on the user's face or hand to recognize hand-over-face gestures.
Instead, we propose a novel sensing configuration on the shoulders, allowing electrodes to remain invisible to both the user and outside observers.
iFace can extract hand-over-face gesture information using shoulder-to-shoulder impedance variations
and achieves an average macro F1 score of 82.58 \% for six common gestures by user-dependent model.
Our results highlight that impedance-based recognition of these nonverbal conversational cues is feasible and can contribute to a smoother user experience with digital devices. 
We present potential applications of iFace to demonstrate how the technology integrates seamlessly into everyday scenarios.

To this end, we make the following \textbf{contributions}:

\begin{enumerate}
    \item We developed iFace, an impedance-based, wearable device for hand-over-face gesture detection and confirmed its feasibility in an experiment with eight participants and six common hand-over-face gestures.
    \item We highlight potential applications of iFace, in particular as an implicit interaction interface for non-verbal communication.
    
\end{enumerate}

\section{Related Work}

Hand-over-face gestures are a subset of body language and can convey various emotions and reactions \cite{Abril07}. 
This section focuses on the related work about hand-over-face gesture recognition.
Existing work for hand-over-face gesture recognition can be grouped into computer vision-based and sensor-based approaches.
For example, in \cite{Dtouch}, an infrared camera (LeapMotion) mounted on the neck takes pictures of the user's face and recognizes the face zones (eyes, nose, and mouth) touch positions with 92\% accuracy. 
Researchers in \cite{Facesight} proposed an infrared camera fixed on a glasses' nose bridge to detect hand-to-face gestures. 
The IR camera was looking downward to capture lower face touching. 
With an accuracy of more than 90\%, the system could detect the fingers touching five areas of the lower face (nose, mouth, chin, and left/right cheeks).
The work \cite{loorak2019hand} presented InterFace to recognize different hand-over-face gestures by smart phone front camera, offering extra possibility of interaction with the phone.
These above solutions often require capturing images of the person to extract the features, which often can achieve high accuracy given high-quality images for facial gesture recognition tasks with deep learning models.
However, for sustained usage in the real world, vision-based methods are often influenced by image quality (e.g., lighting conditions, stability), usability (e.g. requires users to photograph their face), and private issues (esp. for automatic photo-taking methods),
These disadvantages of the computer vision-based approach could be better addressed by the sensor-based methods.
In \cite{Privatetalk}, the authors used the audio signal difference between two earphones (left/right) to detect the hand-to-mouth gesture and trigger a voice input system.
In \cite{Earbuddy}, another earphone-based idea captures the sound when a person’s finger touches their face, thus recognizing hand-to-face gestures, including tapping, double-tapping, and swiping.
A discreet gesture interaction using smart glasses is proposed in \cite{Itchynose}. 
Electrooculography (EOG) was selected as the sensing modality and the authors focused on three gestures with reference to the nose: flicking, pushing, and rubbing with 90\% accuracy. 
In \cite{CheekInput}, photo-reflective sensors were set onto a head-mounted display. 
Their system can recognize the skin deformation caused by the hands touching the cheeks (pushing the face up/down and left/right) with an accuracy higher than 70\%.
Another photo-reflective-based smart glasses is in \cite{Facerubbing} to detect facial rubbing gestures with an accuracy of 97\%. 
In addition, the electromyography (EMG) and capacitive sensing based to identify the body position one touches was demonstrated in \cite{matthies2015botential}.
The solutions discussed above are mostly related to discrete hand-over-face gesture recognition, for example, tapping and double-tapping.
More complex touching face gestures that involved contextual information (e.g., boredom, interest) were not supported.
Additionally, most existing work requires putting sensing units near the face, which is potentially uncomfortable and has social acceptance issues.
In this work, we present iFace, an unobtrusive alternative system to detect hand-over-face gestures to address the challenges.

\section{Design Choices for In-Face Gestures}
Designing for In-Face Gestures requires striking a balance between sensor characteristics, such as accuracy versus unobtrusiveness, and a comprehensive set of meaningful gestures.

\subsection{Choice of Sensor Modality}
The most common sensor-based hand-over-face solution is based on the IMU sensor, as the IMU with a compact package can be easily integrated into the smartwatch and monitor the movement of the hands, for example, food ingestion activity recognition \cite{anderez2018hierarchical}. 
However, the fine hand-over-face gestures cannot be recognized by a single IMU modality, for example, hand-face touch detection and touch area recognition.
Impedance sensing modality leveraging the conductive property of the body provides an opportunity to recognize these fine hand-over-face gestures by attaching each electrode to each shoulder as the touch between hand and face can cause impedance variation, whose magnitude is determined by the touch location and area. 
In addition, the electrical impedance measurement system is more reliable and resistant to noise than other related sensing modalities like Electroencephalogram (EEG), Electromyography (EMG), and capacitive sensing, this is due to the fact that a reference current is injected at one or more specific frequencies, which produces a voltage that is usually easy to measure \cite{bartels2015multi}.
Thus, iFace is designed based on impedance sensing modality.

\subsection{Choice of Sensor Location}
In the realm of hand-over-face gesture recognition, existing sensor placement typically involves positioning sensors in proximity to the hand or head to capture high-quality signals \cite{weng2021facesight,serrano2014exploring}. However, placing sensors close to the face constitutes an intrusive setup, potentially disrupting users' daily activities. While the wrist emerges as an optimal sensor location akin to smartwatches, impedance sensing-based solutions necessitate a closed-loop circuit between two electrodes. This setup would require longer cables if electrodes were attached to both wrists, possibly causing inconvenience to users.
In our study, we navigated a balance between gesture recognition performance and user convenience, opting to place the electrode on the shoulder to monitor impedance variations during hand-over-face gestures. This unobtrusive configuration offers several advantages, for instance, electrodes and cables can be discreetly concealed under clothing, significantly minimizing their impact on individuals' daily routines compared to the location of sensor placements in existing works. 


\subsection{Choice of Gestures}
\begin{figure*}[!t]

\includegraphics[width=1.0\linewidth ]{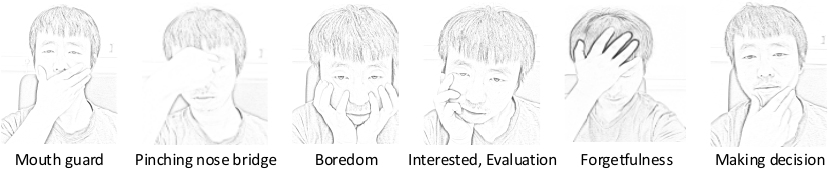}
\caption{Example of hand-over-face gestures}
\label{fig: gesture_list}
\end{figure*}

Hand-over-face gestures are not redundant information; 
they can emphasize the affective cues communicated through facial expressions and speech and give additional information to communication.
In this work, six hand-over-face gestures are selected to be recognized with the use of the proposed iFace, each of the six hand-over-face gestures can convey different potential meanings according to the work \cite{pease2008definitive}, like suspicious, choosing, and thinking. 
These potential meanings conveyed by hand-over-face gestures can not be perceived between each other when the interlocutors are not visible, easily leading to inefficient communication, which can be better addressed by the proposed iFace recognizing such gestures and extracting potential meaning.
Detailed information about these hand-over-face gestures is shown in \cref{tab:gesture_info} and \cref{fig: gesture_list}.

\begin{table*}[!t]
\centering
\footnotesize
\caption{List of hand-over-face gestures, including their description and conveyed meaning according to \cite{pease2008definitive}}
\begin{tabular}{l l l}
\toprule
\textbf{Gesture Name}&\textbf{Description} & \textbf{Potential Meaning} \\
\midrule
Mouth guard & covering mouth using hand & suspicious \\
Pinching the nose bridge & using one's fingers to squeeze or press the area where the nose meets the forehead & skepticism\\
Boredom &resting the head on the two hands, with the face partially covered& disinterest\\
Interested/Evaluation &resting the chin on the fingers while leaning slightly forward& interest or focused attention\\
Forgetfulness & placing one hand on the face, particularly on the forehead or near the eyes& forgetfulness\\
Making decision &resting the chin on the hand or fingers & deciding or choosing\\
\bottomrule
\end{tabular}
\label{tab:gesture_info}
\end{table*}

\section{Implementation}
\subsection{Sensing Principle}
\begin{figure*}[!t]

\includegraphics[width=1.0\linewidth ]{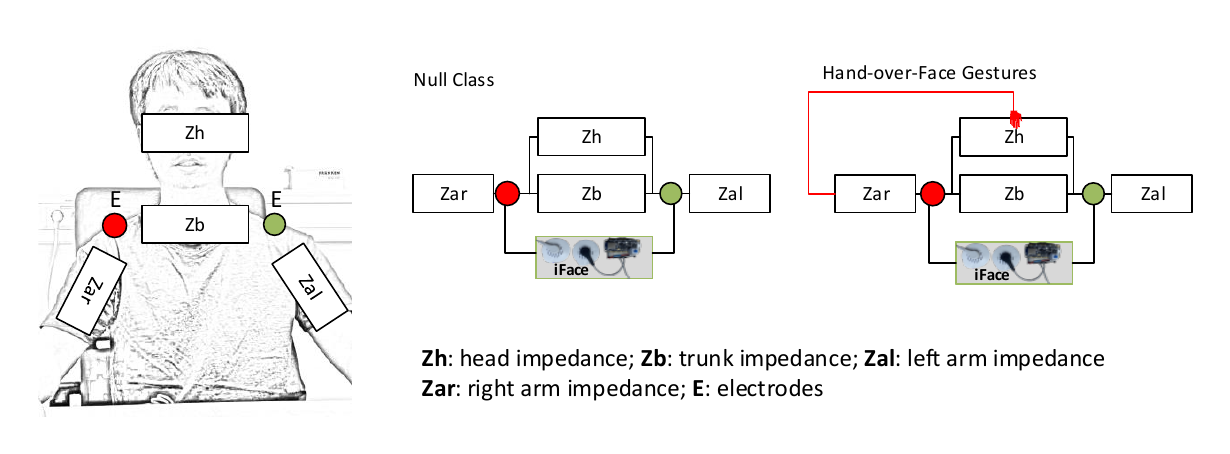}
\caption{Sensing principle of iFace (\textbf{Null class}: the impedance of shoulder-to-shoulder includes only head and trunk part. \textbf{Hand-over-Face gestures}: as the new current path from shoulder to head via arm is built, the arm impedance will be added to the impedance between shoulder-to-shoulder).
Electrodes are hidden under clothing and cannot be observed from the outside.}
\label{fig: sensing_principle}
\end{figure*}

iFace is designed to measure the body impedance variation caused by the hand-over-face gestures based on the fact that the human body containing water is conductive, whose impedance is closely related to the pose and gestures, which are usually discarded signals of motion artefacts of classical bio-impedance measurement \cite{dheman2021towards}; 
\cref{fig: sensing_principle} shows the sensing principle of iFace. 
Two electrodes from the iFace are separately attached to each shoulder to monitor changes in impedance between them.
When there is no hand-to-hand or hand-to-face contact, only one current path connects the shoulders.
However, when users place their hands on their faces, a new current path is created from the shoulder through the hand to the head, leading to a change in impedance between the shoulders.
As the entire body conducts electricity, this impedance variation between shoulders is closely linked to the contact position and area when putting two electrodes to the shoulders separately, enabling the recognition of hand-over-face gestures.
In the future, these electrodes can be seamlessly integrated into clothing, particularly on the shoulders.

\subsection{Hardware Design}
The iFace prototype comprises four integral modules: the analog front-end (AFE) module, the control module, the electrodes, and the power supply.
At the core of the AFE module lies the AD5941 chip by Analog Devices\footnote{https://www.analog.com/en/products/ad5940.html}, chosen for its multifaceted capabilities. 
This chip can generate voltage stimuli in sinusoidal signals, offering a customizable frequency range from 0.015 Hz to 200 kHz. 
Simultaneously, it can measure the response current signal, employing an integrated high-speed transimpedance amplifier to ensure precise current measurements. 
The AD5941 also integrates an FFT hardware accelerator, facilitating the extraction of real and imaginary components from the measurement data.
Driving the AFE is accessed by the nRF52840 controller from Arduino Feather \footnote{https://learn.adafruit.com/adafruit-feather-sense/overview}. 
This component interfaces with the AFE via an SPI bus and facilitates wireless transmission of measurement results to a computer, achieved through Bluetooth interface.
The wet Ag/AgCl electrodes are used to ensure optimal signal acquisition.
For power provision, a compact lithium battery boasting a 500 mAh capacity has been employed in the system design.

\section{System Evaluation}
\subsection{Data Collection}
\begin{figure*}[!t]

\includegraphics[width=1.0\linewidth ]{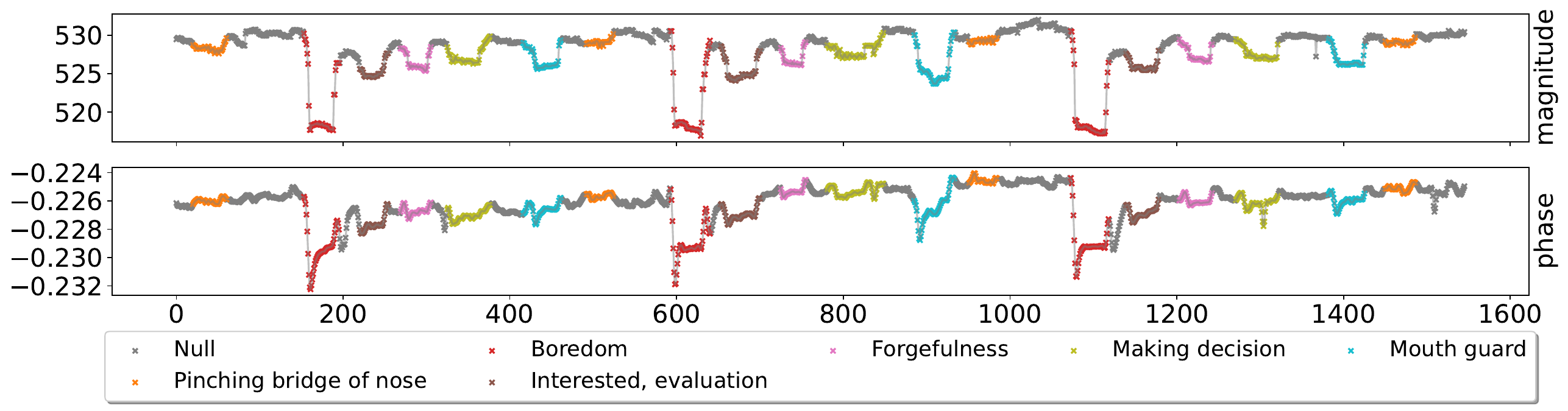}
\caption{An example of raw signals from iFace (sampling rate is 20 Hz)}
\label{fig: raw_signal}

\end{figure*}

To assess the performance of iFace in recognizing hand-over-face gestures, we enlisted eight participants (comprising four females and four males from 22 to 35 years old). 
These participants were asked to perform a set of six hand-over-face gestures while seated at a table, including a null class gesture, as illustrated in \cref{fig: gesture_list}.
Each participant completed four sessions adhering to instructions provided by the conductor, each gesture was performed for around 20 times during each session. Each gesture was maintained for around two seconds.
To facilitate this experiment, we developed a web GUI using JavaScript, enabling us to monitor and record real-time sensor data from iFace via Bluetooth. 
Furthermore, we placed a camera in front of the participants to record video footage to label and validate the collected data.
\cref{fig: raw_signal} shows an example of raw signals from iFace in the experiment.
It can be observed that Boredom gestures caused the most significant magnitude variation, while the impedance variation of the Pinching bridge of the nose was the smallest.
Ethical approval for the study was obtained from the DFKI Ethics Board.

\subsection{Data Processing}
\begin{figure*}[!t]

\includegraphics[width=1.0\linewidth ]{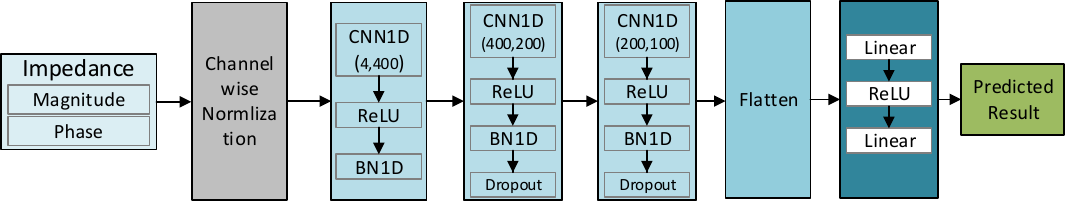}
\caption{Data processing pipeline for hand-over-face gesture recognition. The impedance data includes two channels: magnitude and phase, which are preprocessed by channel-wise normalization method before being inputted to the neural network}
\label{fig: data_processing_pipeline}

\end{figure*}

\cref{fig: data_processing_pipeline} shows the data processing pipeline for hand-over-gesture recognition.
The raw data from iFace has a single sensing modality: impedance data, whose sampling rate is 20 Hz after synchronization.
The impedance data from the AFE module includes magnitude and phase.
A channel-wise normalization was adapted within slide windows before inputting the raw data into the neural network.
Three one-dimensional convolutional layers were used to extract the features from the raw data, followed by two linear layers for feature classification, implemented using PyTorch with its backend.
A grid search method with a window size search space from 50 to 120 was implemented to select an optimal window size of instances inputting to the neural network.
The sliding window was resampled from the raw time series with a step size of one for training processing, while the step size was configured as 30 for test.
The neural network was trained using the cross-entropy loss function and the Adam optimizer with 1e-5 learning rate and  0.9 and 0.999 for $\beta_1$ and $\beta_2$, respectively.
Since the dataset is very imbalanced, the null classes instance is much more than other classes.
A weight parameter calculated from the number of labels present inside the training dataset was added to the Cross-Entropy loss to give more importance to a certain class in the Cross-Entropy loss.
Each model is trained for 200 epochs in this work.

\subsection{Classification Result and Discussion}

\begin{figure*}[!t]
\includegraphics[width=1.0\linewidth ]{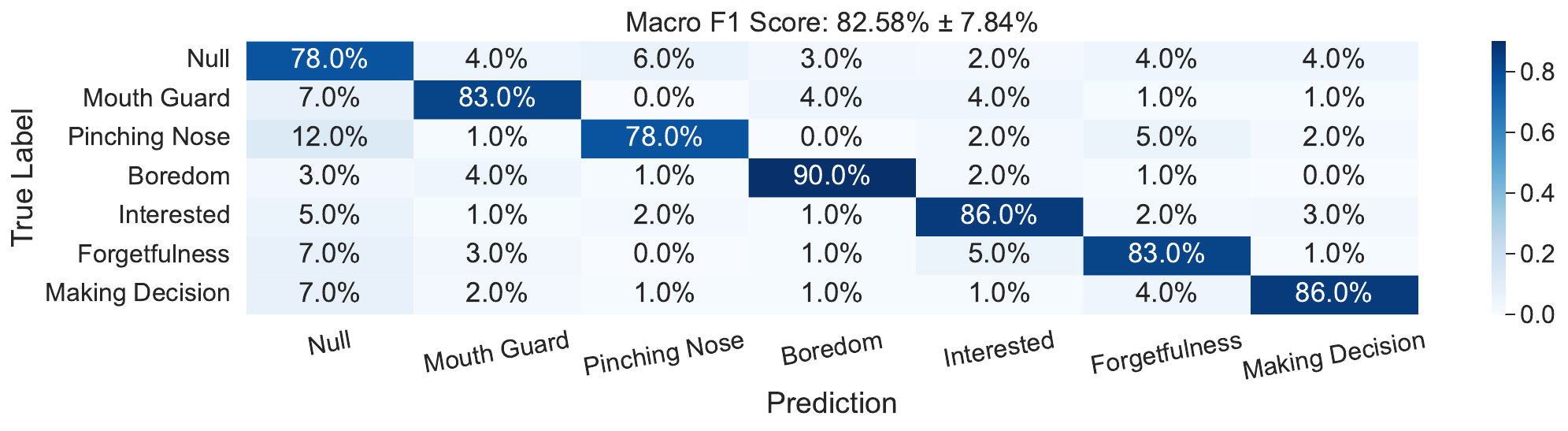}
\caption{Joint confusion matrices for hand-over-face gesture recognition from eight subjects together}
\label{fig:cm-gesture}
\end{figure*}

Since we found the participants with different native body impedance performing the same hand-over-face gestures differently (different dominant hands, different touch ways), a user-dependent model was trained to recognize the activities across the sessions by the leave-one-session-out procedures. 
This study selected the macro F1 score as the metric, computed using the arithmetic mean (unweighted mean) of all the per-class F1 scores. 
Thus, it treats all classes equally.
\cref{fig:cm-gesture} presents the classification result and joint confusion matrices from eight subjects.
The model with the input of impedance sensing modality achieved an average Macro F1 Score of 82.58 \%.
The most confusing classes are the same: Null class and Pinching nose gesture.
Because the contact area between the hand and face of the pinching nose gesture is the smallest, it leads to the impedance variation being minimal compared to other activities, as shown in \cref{fig: raw_signal}, which could be confused with the null class if the contact between the finger and nose bridge is not good.
In contrast, the Boredom gesture is the best recognized of all the gesture classes by the neural network model, with an average recognition recall of 90.0 \%. 
The Boredom gesture is a unique hand-over-face gesture requiring resting the head on the two hands, with the face partially covered.  
The contact area of the boredom gesture is the most significant, and two small current paths through both arms are formed when a subject performs such a gesture, resulting in the most considerable impedance reduction between shoulders. 
Thus, it can be easily distinguished among the six kinds of gestures.

\section{Potential Applications for Impedance-Based Hand-over-Face Gesture Recognition}
Our work demonstrated the performance of iFace for six types of hand-over-face gestures based on shoulder-to-shoulder impedance variation information.
In this section, we aim to provide a vision for future researchers, illustrating how they can build upon our concept and apply it to a broader array of practical applications.

\subsection{Enriching Implicit Interaction Through iFace}
Face-based input has been widely used for many scenarios, like mobile interaction, auto-screen rotation, and authentication \cite{cheng2012irotate,de2015feel,zhao2016augmenting}.
However, most of these applications are computer vision-based. 
iFace proposes an impedance sensing-based solution for face-based input. 
Thus, it leverages several advantages, such as being privacy-protecting, lightweight, and unobtrusive. 
Unlike existing face-based touch input, like tapping and
flicking, iFace can provide diverse touch patterns based on hand-over-face gestures, expanding interaction possibilities.
In addition, the touch input in existing works \cite{loorak2019hand} is designed as an intentional behavior, and the users need to learn the operating instructions explicitly.
Yet, hand-over-face gestures often encompass body language as well, such as cues like boredom and interest. 
They are often performed by people unconsciously reacting to their surroundings. 
Such involuntary hand-over-face gestures detected by iFace can be used to enrich implicit interactions with digital devices.
For example, when users watch a video, the computer automatically changes the channel if they perform an involuntary boredom hand-over-face gesture.

\subsection{Augmenting Non-visual Communication}
It is challenging for conversation partners to convey interpersonal cues in non-visual communication because the interlocutors cannot see facial expressions and body language.
\citet{vermeulen2016heartefacts} have demonstrated that technology can mediate this interaction, e.g., HeartChat \cite{hassib2017heartchat} uses heart-rate-augmented mobile messaging to support empathy and awareness.
Other physiological sensing modalities, like blood volume pulse, galvanic skin response, and electroencephalography, are also explored to extract the emotional cues and address this challenge \cite{khan2016recognizing,westerink2008computing,ferdinando2014emotion}.
With iFace, we propose using the body impedance sensing modality to recognize hand-over-face gestures for implicit detection of bodily cues. 
We envision several potential application scenarios based on iFace.
like using hand-over-face gestures in online meetings to convey real-time bodily cues when video quality is poor. It can also integrate with text, sending detected gestures and text to convey interpersonal cues while typing messages.

\subsection{Understanding Cognitive States}
Spontaneous self-touches and self-grooming gestures are believed to have connections with thought formulation, information processing, and emotion regulation. \cite{freedman1977hands}.
There is empirical evidence that some of these hand-over-face gestures serve as cues for recognizing cognitive states \cite{mahmoud2016automatic}.
iFace also has the potential to help understand the cognitive mental state by detecting specific hand-over-face gestures related to cognitive processes. 
Thus, iFace can provide an additional sensing modality for multi-modal inference of a user's cognitive state in many application scenarios. 
For example, amplifying the audience-performer connection \cite{sugawa2021boiling} and sensing implicit audience engagement \cite{hassib2017engagemeter}.

\section{Conclusion}
In this work, we presented iFace, an unobtrusive, wearable impedance-sensing device for hand-over-face gesture recognition. Firstly, we proposed a general concept of hand-over-face gesture detection based on body part impedance variation caused by the hand-face interaction. 
Then, we designed and implemented an unobtrusive, wearable impedance-sensing device for body impedance measurement. 
Additionally, we demonstrated iFace's performance in classifying six hand-over-face gestures performed by eight participants using a lightweight neural network model. 
The model achieved an average Macro F1 Score of 82.58\% with the input of a single impedance sensing modality. 
Finally, we envisioned several potential application scenarios based on iFace, in which users can leverage the recognized hand-over-face gestures to enrich implicit interaction and augment non-visual communication to support empathy and awareness. 

\section*{Acknowledgements (not compulsory)}

This work has been supported by the BMBF (German Federal Ministry of Education and Research) in the project SocialWear (01IW20002).

\bibliographystyle{ACM-Reference-Format}
\bibliography{sample-base}

\end{document}